\documentclass[twocolumn,showpacs,preprintnumbers,prl,aps,amssymb,superscriptaddress]{revtex4}
\usepackage{graphicx}
\usepackage{dcolumn}
\usepackage{bm}
\usepackage{color}

\begin{document}

\newcommand{\ie}{{\it i.e.}}
\newcommand{\eg}{{\it e.g.}}
\newcommand{\etal}{{\it et al.}}


\title{ Quasiparticle Heat Transport in Ba$_{1-x}$K$_x$Fe$_2$As$_2$ : 
Evidence for a $k$-dependent Superconducting Gap without Nodes }


\author{X. G. Luo}
\affiliation{D\'epartement de physique
\& RQMP, Universit\'e de Sherbrooke, Sherbrooke, Canada}

\author{M.~A.~Tanatar}
\altaffiliation{e-mail tanatar@ameslab.gov} 
\affiliation{Ames Laboratory, Ames, Iowa 50011, USA}

\author{J.-Ph. Reid}
\affiliation{D\'epartement de physique
\& RQMP, Universit\'e de Sherbrooke, Sherbrooke, Canada}

\author{H. Shakeripour}
\affiliation{D\'epartement de physique
\& RQMP, Universit\'e de Sherbrooke, Sherbrooke, Canada}

\author{N. Doiron-Leyraud } \affiliation{D\'epartement de physique
\& RQMP, Universit\'e de Sherbrooke, Sherbrooke, Canada}

\author{N.~Ni}
\affiliation{Ames Laboratory, Ames, Iowa 50011, USA}
\affiliation{Department of Physics and Astronomy, Iowa State
University, Ames, Iowa 50011, USA }

\author{S.~L.~Bud'ko}
\affiliation{Ames Laboratory, Ames, Iowa 50011, USA}
\affiliation{Department of Physics and Astronomy, Iowa State
University, Ames, Iowa 50011, USA }

\author{P.~C.~Canfield}
\affiliation{Ames Laboratory, Ames, Iowa 50011, USA}
\affiliation{Department of Physics and Astronomy, Iowa State
University, Ames, Iowa 50011, USA }

\author{Huiqian~Luo}
\affiliation{ National Laboratory for Superconductivity, Institute of Physics and
National Laboratory for Condensed Matter Physics, P. O. Box 603 Beijing, 100190,
P. R. China }

\author{Zhaosheng Wang}
\affiliation{ National Laboratory for Superconductivity, Institute of Physics and
National Laboratory for Condensed Matter Physics, P. O. Box 603 Beijing, 100190,
P. R. China }

\author{Hai-Hu Wen}
\affiliation{ National Laboratory for Superconductivity, Institute of Physics and
National Laboratory for Condensed Matter Physics, P. O. Box 603 Beijing, 100190,
P. R. China }
\affiliation{Canadian Institute for Advanced Research, Toronto, Ontario, Canada}

\author{Ruslan Prozorov}
 \affiliation{Ames Laboratory, Ames,
Iowa 50011, USA} \affiliation{Department of Physics and Astronomy,
Iowa State University, Ames, Iowa 50011, USA }

\author{Louis Taillefer}
\altaffiliation{e-mail louis.taillefer@physique.usherbrooke.ca }
\affiliation{D\'epartement de physique \& RQMP, Universit\'e de
Sherbrooke, Sherbrooke, Canada} \affiliation{Canadian Institute for
Advanced Research, Toronto, Ontario, Canada}

\date{\today}


\begin{abstract}
The thermal conductivity $\kappa$ of the iron-arsenide superconductor Ba$_{1-x}$K$_x$Fe$_2$As$_2$ ($T_c \simeq$ 30~K)
was measured in single crystals at temperatures down to $T \simeq 50$~mK ($\simeq T_c$/600) and in magnetic fields up to $H = 15$~T ($\simeq H_{c2}$/4).
A negligible residual linear term in $\kappa/T$ as $T \to 0$ shows that there are no zero-energy quasiparticles in the superconducting state.
This rules out the existence of line and in-plane point nodes in the superconducting gap, imposing strong constraints on the symmetry of the order parameter. It excludes $d$-wave symmetry, drawing a clear distinction between these superconductors and the high-$T_c$ cuprates.
However, the fact that a magnetic field much smaller than $H_{c2}$ can induce a residual linear term indicates that the gap must be very small on part of the Fermi surface, whether from strong anisotropy or band dependence, or both. 

\end{abstract}


\maketitle


Several experiments have been performed to address the symmetry of the
order parameter in iron arsenide superconductors.
Early studies on mostly polycrystalline samples of $R$FeAs(O,F), where $R$ = La, Nd or Sm,
have led to indications that appear contradictory.
While point-contact spectroscopy \cite{Tesanovic},
ARPES \cite{Adam} and penetration depth \cite{Martin,Matsuda,Carrington} point to a full superconducting
gap without nodes, specific heat \cite{heat_unconventional} and NMR \cite{NMR} data
were interpreted in terms of a nodal superconducting gap.
Reports on single crystals of doped BaFe$_2$As$_2$,
also appear contradictory.
In (Ba,K)Fe$_2$As$_2$, ARPES studies have found an isotropic superconducting gap with a magnitude
of 12 meV on one Fermi surface and 6 meV on another \cite{Adam122,ARPESchinese,Ding}, specific heat measurements are broadly consistent with an $s$-wave gap of 6 meV \cite{heat_cap_BaK}, as are muon
measurements of the superfluid density \cite{Hiraishi}.
By contrast, penetration depth studies find a power-law variation \cite{MartinBaK}, as in Co-doped BaFe$_2$As$_2$ \cite{Gordon}, as opposed to the
exponential temperature dependence expected of an isotropic $s$-wave gap. 

In an attempt to shed further light on the structure of the superconducting gap,
we have measured the thermal conductivity $\kappa$ of K-doped BaFe$_2$As$_2$.
Heat transport is a powerful probe of symmetry-imposed nodes in the superconducting gap \cite{NJP2009}.
We find a negligible residual linear term $\kappa_0/T$ in $\kappa/T$ as $T \to 0$, strong evidence that there are no nodes in the gap of this superconductor.
Indeed, a line of nodes would have given a sizable and universal (i.e. impurity independent) $\kappa_0/T$ \cite{NJP2009}, as in cuprates \cite{Taillefer1997}, 
ruthenates \cite{Suzuki2001} and some heavy-fermion superconductors \cite{Shakeripour2009}.
Given the large impurity scattering rate in our samples, point nodes would also have given a sizable (albeit non-universal) $\kappa_0/T$ \cite{NJP2009} -- 
unless they happend to lie along the $c$-axis, perpendicular to the direction of heat flow in our measurements.
However, a magnetic field $H$ applied along the $c$-axis induces a finite $\kappa_0/T$ even when $H << H_{c2}$. This shows that the superconducting gap 
must be very small on some part of the Fermi surface, 
either because of a pronounced anisotropy on one Fermi surface (whereby the gap has a deep minimum in some directon) or because
of pronounced band dependence, causing one Fermi surface sheet to have a very small gap.

\begin{figure} [t]
\centering
\includegraphics[width=8.5cm]{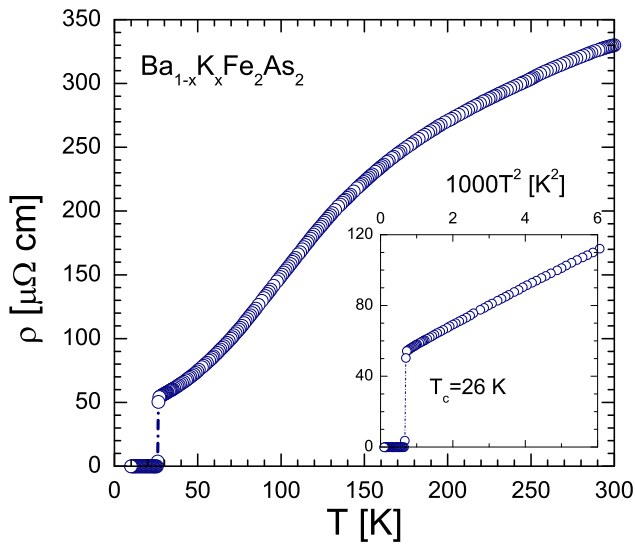}
\caption{(Color online) Temperature dependence of the in-plane
resistivity $\rho(T)$ of Ba$_{1-x}$K$_x$Fe$_2$As$_2$, with $x \simeq 0.25$ and $T_c = 26$~K.
Inset: same data plotted as a function of $T^2$. The line is a linear fit to $\rho(T)= \rho _0 + A T^2$ below 70 K.}
\label{resistivity}
\end{figure}


Single crystals of Ba$_{1-x}$K$_x$Fe$_2$As$_2$ were grown from FeAs flux \cite{Wen_crystals}. The doping level of the two samples used in the present
study, labeled A and B,
was determined from the $c$-axis lattice parameter,
giving $x$= 0.25 for sample A ($T_c$ = 26~K) and $x$= 0.28 for sample B ($T_c$ = 30~K), both on the underdoped side,
{\it i.e.} below optimal doping ($x \simeq$ 0.4) \cite{Rotter2}.

Samples were cleaved into rectangular
bars with typical size 1.5 $\times$ 0.3 $\times$ 0.05~mm$^3$. Silver
wires were attached to the samples with a silver-based alloy,
providing ultra-low contact resistance of the order of 100 $\mu
\Omega$. Thermal conductivity was measured along the [100] direction in the
tetragonal crystallographic plane in a standard
one-heater-two thermometer technique \cite{Hawthorn2007}.
The magnetic field $H$ was applied along the [001] tetragonal axis. All
measurements were done on warming after cooling in constant $H$ from above $T_c$ to ensure a homogeneous field distribution
in the sample.
All aspects of the charge and heat transport are qualitatively the same in both samples, with minor quantitative differences. 
For simplicity, only the data for sample A is displayed here. 
Whenever quoted, we provide quantitative values for both samples.

In Fig.~\ref{resistivity}, we show the electrical resistivity $\rho(T)$ of sample A as a function of temperature.
Below $\sim$150~K, $\rho(T)$ shows a notable downturn, followed by a range where $\rho(T)$ is well described
either by a power-law dependence, $\rho(T)= \rho _0 + A T^n$, with $n$ between 1.6 to 1.8, or by $\rho(T)= \rho _0 + A T^2$,
below 70 K or so.
Using the latter fit, we get $\rho _0$ = 47 $\mu \Omega$~cm (sample A) and 28 $\mu \Omega$~cm (sample B).
Since the same contacts are used for electrical and thermal transport, 
this allows us to accurately estimate the normal-state thermal conductivity $\kappa_{\rm N}/T$ in the $T \to 0$ limit, via the
Wiedemann-Franz law, $\kappa_{\rm N}/T = L_0 / \rho_0$ where $L_0 \equiv \frac{\pi^2}{3} (\frac{k_{\rm B}}{e})^2$,
giving $\kappa_{\rm N}/T$ = 520 $\mu$W/K$^2$ cm (sample A) and 875 $\mu$W/K$^2$ cm (sample B).

\begin{figure} 
\centering
\includegraphics[width=8.5cm]{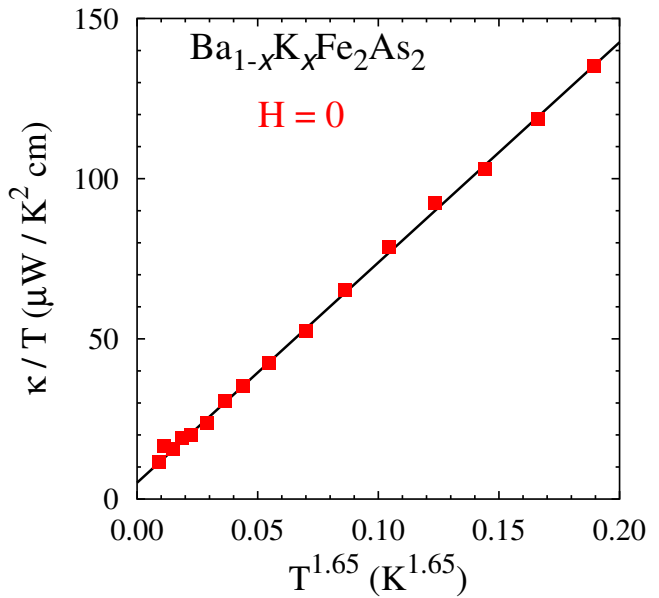}
\caption{(Color online) Temperature dependence of the thermal conductivity $\kappa(T)$, plotted as $\kappa/T$ vs $T^{1.65}$, in zero magnetic field. The line is a linear fit over the temperature range shown.}
\label{beta}
\end{figure}


{\it Residual linear term in zero magnetic field}.
The thermal conductivity $\kappa(T)$ of sample A, measured for $H$=0, is shown in Fig.~\ref{beta}.
The data is plotted as $\kappa/T$ vs $T^{1.65}$.
The linear fit displayed in Fig.~\ref{beta} shows that the data below 0.4 K is well described by the function $\kappa/T = a + b T^{\alpha}$,
with $a \equiv \kappa_0/T$ = 5 $\mu$W / K$^2$ cm and $\alpha = 1.65$.
This same function describes the data of sample B equally well, with $a \equiv \kappa_0/T$ = 7 $\mu$W / K$^2$ cm and $\alpha = 1.5$.
The first term is the residual linear term of interest here \cite{NJP2009}. 
The second term is due to phonons, which at low temperatures are scattered by the sample boundaries.
Although the latter term would be expected to give $\kappa_p \propto T^3$, {\it i.e.} $\alpha = 2$, measurements on single crystals
with smooth surfaces rarely show this textbook behavior because of a specular reflection off the surface, and in practice $1 < \alpha < 2$ 
\cite{Li2008,Sutherland2003,Li2007}.

The magnitude of the residual linear term extracted from the fits in Fig.~\ref{beta} is extremely small.
In similar measurements on samples where no residual linear term is expected, $\kappa_0/T$ was indeed found to be zero, within an error bar of $\pm 5$ 
$\mu$W / K$^2$ cm or so \cite{Li2008,Boaknin2003}.
Within those error bars, our two samples of Ba$_{1-x}$K$_x$Fe$_2$As$_2$ exhibit negligible residual linear terms.
Let us put these minute $\kappa_0/T$ values into perspective. Comparison with the normal-state conductivity, gives the ratio $(\kappa_0/T) / (\kappa_{\rm N}/T) \simeq 1 \%$ in both samples.
Secondly, these $\kappa_0/T$ values are much smaller than theoretical expectations for a nodal superconductor.
(For a gap without nodes, $\kappa_0/T$ should be zero \cite{NJP2009}.)
For a quasi-2D $d$-wave superconductor, with four line nodes along the $c$-axis, the residual linear term is given, in the clean limit ($\hbar \Gamma_0 << \Delta_0$), 
by \cite{Graf1996,Durst2000,Hawthorn2007,NJP2009}
$\kappa_{0}/T = (k^2_{\rm B}/6 c) (k_{F} v_{F} / \Delta_{0})$,
where $c$ is the interlayer separation, $k_F$ and $v_F$ the 
Fermi wavevector and velocity at the node, respectively, and $\Delta_0$ the gap maximum.
Taking $c$ = 6.6~$\AA$, $v_F$ = 0.3~eV~$\AA$ = $0.45 \times 10^5$ m/s \cite{Kaminski}, and a Fermi wavevector $k_F$ = 0.35~$\times \pi/a$ = 0.28~\AA$^{-1}$, and assuming a weak-coupling $\Delta_0 = 2.14$ $k_{\rm B} T_c$, we get $\kappa_{0}/T$ = 140 $\mu$W / K$^2$ cm. 
This is at least 20 times larger than the value extracted from our fits to the data.
In those materials where universal heat
transport has been verified, the measured value of $\kappa_0/T$ is
in good quantitative agreement with this theoretical expectation
\cite{Hawthorn2007,Suzuki2001,Shakeripour2009}. 
Thus we can safely conclude that the gap in Ba$_{1-x}$K$_x$Fe$_2$As$_2$ does not contain a line of nodes anywhere on the Fermi surface. In particular, this rules out $d$-wave symmetry, whether $d_{x^2 - y^2}$ or $d_{xy}$. This result sets cuprates and iron arsenides apart as two distinct types of high-temperature superconductors.

It is in principle possible for the superconducting gap in pnictides to have point nodes as opposed to line nodes, one of the scenarios suggested by penetration depth studies \cite{Gordon}.
The zero-energy quasiparticles associated with point nodes give a residual linear term which grows with impurity scattering \cite{Graf1996}, so that $\kappa_{0}/T$ can become a substantial fraction of $\kappa_{\rm N}/T$ \cite{Graf1996}.
Given $\rho_0$, we can estimate the normal-state impurity scattering rate $\Gamma_0$ roughly from the plasma frequency $\omega_{\rm p} = c / \lambda_0$,
where $\lambda_0$ is the penetration depth, approximately equal to 200 nm \cite{Gordon}. This gives $\Gamma_0 / T_c \simeq 1.7$ and 0.9 for sample A and B, respectively.
This is very substantial, and would give a large residual linear term, comparable to the case of the line node.
We therefore conclude that point nodes in the gap of Ba$_{1-x}$K$_x$Fe$_2$As$_2$ are also unlikely, unless they are located along the $c$-axis and do not contribute to in-plane transport.
Needless to say, our data also rule out the possibility of an entirely gapless (or ungapped) Fermi surface, proposed by some authors \cite{Stanescu},
at least down to the 1 \% level.

The nodes we have discussed so far are imposed by symmetry, the result of a sign change in the order parameter around the Fermi surface. Such symmetry-related nodes are broadened by impurity scattering, giving rise to a sizable $\kappa_{0}/T$.
The superconducting gap can also go to zero in certain directions because of a pronounced anisotropy that is not imposed by symmetry. 
However, such "accidental" nodes in a gap with $s$-wave symmetry will be lifted by impurity scattering, making the gap more isotropic \cite{Hirschfeld}. Our zero-field data is  consistent with an $s$-wave gap, including one with strong anisotropy.


\begin{figure}[t]
\includegraphics[width=9cm]{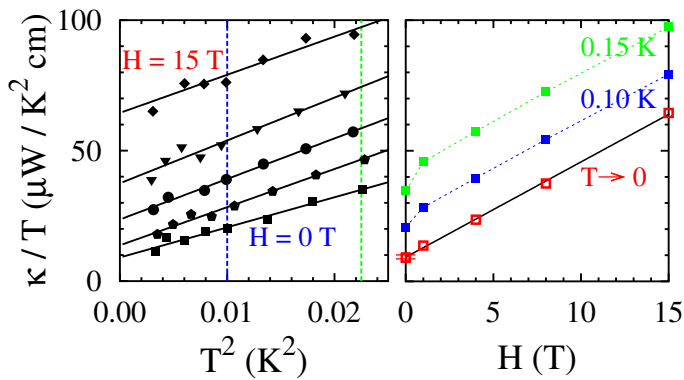}
\caption{(Color online)  Left panel: Temperature dependence of the
thermal conductivity measured in magnetic fields
of 0, 1, 4, 10 and 15~T (from bottom to top), plotted as
$\kappa/T$ vs $T^2$. Solid lines are a linear fit to each curve in
the range shown. Vertical dashed lines indicate $T = 0.1$~K (blue)
and $T = 0.15$~K (green). The value of $\kappa/T$ at those two
temperatures is plotted vs magnetic field in the right panel.
Right panel: isotherms of $\kappa/T$ as a function of magnetic field
$H$, for $T \to 0$ (obtained by extrapolating the linear fits in the
left panel to $T=0$), 0.1 K and 0.15 K. In all three cases,
$\kappa/T$ rises approximately linearly with $H$, with the same
slope. The solid line is a linear fit to the $T \to 0$ data, also
reproduced in Fig.~\ref{comparison}.}
\label{T2field}
\end{figure}

\begin{figure}[tb]
\includegraphics[width=8.5cm]{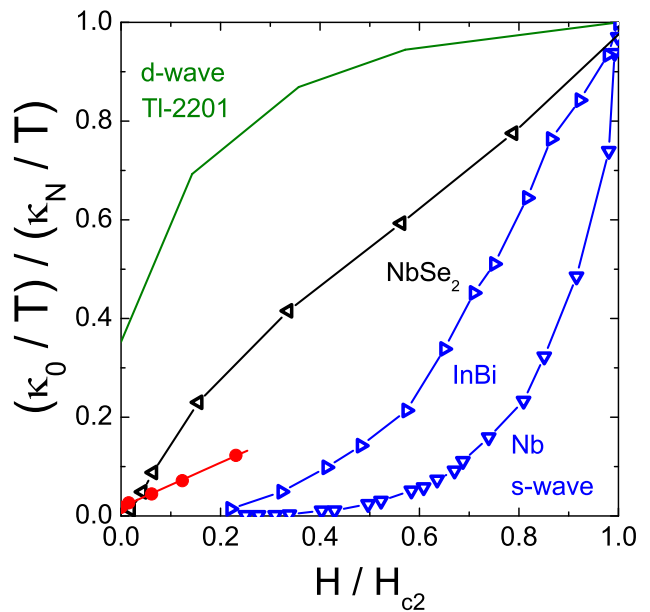}
\caption{(Color online) Residual linear term $\kappa_0/T$ in the thermal conductivity of various superconductors as a function of magnetic field $H$, presented on normalized scales $\kappa_s/\kappa_N$ vs $H/H_{c2}$.
For Ba$_{1-x}$K$_x$Fe$_2$As$_2$ (red circles), we use $\kappa_N/T$ obtained from the Wiedemann-Franz law applied to the extrapolated residual resistivity
$\rho_0$ (see text) and $H_{c2}$=65~T \cite{Upper_critical_field}.
The data for the clean and dirty isotropic $s$-wave superconductors Nb and InBi, respectively, are reproduced from \cite{Li2007}.
The data for the $d$-wave superconductor Tl-2201 is for a strongly overdoped sample ($T_c = 15$~K and $H_{c2} \simeq 7$~T) \cite{Proust2002}.
The data for multi-band superconductor NbSe$_2$ ($T_c = 7$~K and $H_{c2} = 4.5$~T) is from \cite{Boaknin2003}.
}
\label{comparison}

\end{figure}

{\it Field dependence of thermal conductivity}.
The effect of a magnetic field $H$ ($H \parallel c$) on the thermal conductivity of Ba$_{1-x}$K$_x$Fe$_2$As$_2$ is displayed in Fig.~\ref{T2field}.
In the panel on the left, $\kappa/T$ curves are seen to shift upwards almost rigidly with field.
In the right panel, the value of $\kappa/T$ at three temperatures ($T \to 0$, $T = 0.1$~K and $T = 0.15$~K) is seen to rise linearly with $H$ up to our highest field of 15~T, which corresponds roughly to $H_{c2}/4$ (using $H_{c2}$ at $T \to 0$ as approximately 65~T \cite{Upper_critical_field}).
In Fig.~\ref{comparison}, we compare the field dependence of $\kappa_0/T$ in Ba$_{1-x}$K$_x$Fe$_2$As$_2$ with the dependence in various
other superconductors, using normalized conductivity and field scales, $\kappa_s/\kappa_N$ and $H/H_{c2}$. 
In a $d$-wave superconductor like the overdoped cuprate Tl-2201 ($T_c = 15$~K and $H_{c2} \simeq 7$~T), $\kappa_0/T$ rises very steeply at the lowest fields \cite{NJP2009,Hawthorn2007,Proust2002}, 
roughly as $\kappa_0/T \propto \sqrt{H}$, following the density of delocalized zero-energy quasiparticles outside the vortex cores \cite{Hirschfeld-vortex}.
By contrast, in an isotropic $s$-wave superconductor like Nb, the rise in $\kappa_0/T$ is exponentially slow at low fields as it relies on the tunneling of quasiparticles between localized states inside adjacent vortex cores, which at low fields are far apart.
In many materials, however, the situation is not so clearcut. A good example are the multi-band superconductors MgB$_2$ \cite{MgB2} and NbSe$_2$ \cite{Boaknin2003}. Here the magnitude of the $s$-wave superconducting gap is very different on two sheets of the Fermi surface.
In both materials, the small gap is roughly one third of the large gap, so that a field $H \simeq H_{c2}/9$ is sufficient to kill superconductivity on the small-gap Fermi surface, which can then contribute its full normal-state conductivity even deep inside the vortex state.
Specifically, at $H = H_{c2}/5$, $\kappa_0/T$ is already half (one third) of $\kappa_N/T$ in MgB$_2$ (NbSe$_2$). It is one tenth in Ba$_{1-x}$K$_x$Fe$_2$As$_2$.
By comparison, $\kappa_0/T$ is still negligible in a single-gap superconductor like pure Nb or disordered InBi (see Fig.~\ref{comparison}).
This shows that the superconducting gap must be small on some part of the Fermi surface of Ba$_{1-x}$K$_x$Fe$_2$As$_2$, relative to the gap maximum which controls $H_{c2}$.

In Fig.~\ref{comparison}, we show data for
NbSe$_2$ \cite{Boaknin2003}, where we see that
$\kappa_0/T = 0$ at $H=0$ and $\kappa_0/T$ rises linearly above $H_{c2}/30$, with a slope of 1.67 in the normalized units
of Fig.~\ref{comparison}.
In Ba$_{1-x}$K$_x$Fe$_2$As$_2$, $\kappa_0/T$ also rises linearly, with a normalized slope of 0.46 in sample A and 0.21 in sample B.
In a multi-band scenario, the magnitude of this slope is roughly proportional to the value of the normal-state conductivity $\kappa_N/T$ of the small-gap Fermi surface relative to the overall conductivity. The fact that the slope in absolute units is larger in sample A (3.7 $\mu$W / K$^2$~cm~T) than in sample B (2.8 $\mu$W / K$^2$~cm~T) eventhough its total normal-state conductivity $\kappa_N/T = L_0/\rho_0$ is {\it smaller} (520 vs 875 $\mu$W/K$^2$~cm) is suggestive of a multi-band situation with the impurity scattering rate being different on different Fermi surfaces.

A recent heat transport study of the low-$T_c$ nickel-arsenide superconductor BaNi$_2$As$_2$ ($T_c = 0.7$~K) \cite{Kurita} gave $\kappa_0/T = 0$, as here in Ba$_{1-x}$K$_x$Fe$_2$As$_2$ ($T_c = 26-30$~K).
However, it found a much slower increase of $\kappa_0/T$ at low $H/H_{c2}$, consistent with a gap that is large everywhere on the Fermi surface.


We conclude that there are no nodes in the superconducting gap of Ba$_{1-x}$K$_x$Fe$_2$As$_2$, at least at $x \simeq 0.25-0.28$, with the possible exception of point nodes along the $c$-axis. 
This excludes $d$-wave symmetry, and any other symmetry that requires line nodes on the multi-sheet Fermi surface of this superconductor.
Symmetries consistent with this constraint include $s$-wave and $s_{\pm}$, whereby a full gap changes sign from the electron Fermi surface to the hole Fermi surface \cite{s+-symmetry}. 
From the rapid rise of $\kappa_0/T$ with magnetic field at very low fields, we infer that the gap must be very small on some portion of the Fermi surface.
This $k$-dependence of the gap magnitude can come from angle dependence or band dependence, or both.
In many experiments, the presence of a very small gap could mimic that of a node.
%


M.A.T. acknowledges continuing cross-appointment with the Institute of Surface Chemistry, N. A. S. of Ukraine. 
Work at the Ames Laboratory was supported by the Department of Energy-Basic Energy Sciences under Contract No. DE-AC02-07CH11358.
R.P. acknowledges support from the Alfred P. Sloan Foundation.
L.T. acknowledges support from the Canadian Institute for Advanced Research, a Canada Research Chair, NSERC, CFI and FQRNT.


\end{document}